\documentclass[aps,prl,twocolumn,twoside,nolongbibliography, notitlepage, preprintnumbers,showpacs,  showkeys,byrevtex,floatfix,amsmath,amssymb]{revtex4-2}

\usepackage{marginnote,xparse,changepage,dcolumn,amsfonts, array,graphicx,mathtools,multirow,placeins,dsfont}

\usepackage[T1]{fontenc} 
\usepackage{epsfig}
\usepackage{float}
\usepackage{hyperref}
\usepackage{tikz}

\newcommand{\ZZ}{\mathbb{Z}}

\begin{document}

\title{Compatibility of Braiding and Fusion on Wire Networks}

\author{A. Conlon$^{1,2}$}
\email[Electronic address: ]{aaron@stp.dias.ie}
\author{J. K. Slingerland$^{1,2}$}
\email{Electronic address: joost@thphys.nuim.ie}
\affiliation{$~^{1}\,$Department of Theoretical Physics, Maynooth University, Ireland.}
\affiliation{$~^{2}\,$School of Theoretical Physics, Dublin Institute for Advanced Studies,
10 Burlington Road, Dublin 4, Ireland}
\date{\today}

\begin{abstract}
\noindent  
  Exchanging particles on graphs, or more concretely on networks of quantum wires, has been proposed as a means to perform fault tolerant quantum computation. This was inspired by braiding of anyons in planar systems. 
 However, exchanges on a graph are not governed by the usual braid group but instead by a graph braid group. 
 By imposing compatibility of graph braiding with fusion of topological charges, we obtain generalized hexagon equations.
 We find the usual planar anyons solutions but also more general braid actions. We illustrate this with Abelian, Fibonacci and Ising fusion rules.
\end{abstract}

\preprint{DIAS-STP-22-01}

\maketitle

Two decades ago, it was realized that an inherently fault tolerant quantum computation scheme could be implemented using the exchange statistics of anyons, quasiparticles in planar systems.
This gave birth to the field of \textit{topological quantum computation} (TQC), \cite{Kitaev2003, Kitaev2006, Nayak2008}.
Physical systems that can host anyons include fractional quantum Hall (FQH) states \cite{TopQubits_from_FQHE}.
In fact, the Aharonov-Bohm signature of Abelian anyon exchange statistics in FQH states was recently directly observed \cite{FQHE_nu_OneThird_AbelianStat,2020AnyonObservation}.
Much of the effort in TQC has recently focused on one-dimensional systems since it was proposed \cite{Alicea_Junction} that the braiding of anyon-like excitations could be performed at junctions in networks of semiconductor wires.
The possibility of transporting Majorana modes around networks has been extensively investigated, \cite{Aaron2019,LuukMachineLearning,Helical_MBS_Refael, MajFerm_TopPhase,UniversalQC_Semiconductor,ScheurerShnirnman,Boosting,BraidingEffects_Anyon_TopSupercond, Topological_Josephson_MBS,Major_PianoKey,BraidingEffects_Anyon_TopSupercond}. 
Exchanging particles on networks, or graphs, rather than in the plane, merits closer investigation.
The exchange statistics of $N$ identical particles are governed by the representations of the fundamental group of the configuration space of the system. 
This is the space of unordered collections of $N$ distinct particle positions in the relevant geometry ~\cite{Leinaas_Myrheim,SuperSelection_Braid_Schroer,Frohlich_Local_QFT}.  
For particles on the plane this fundamental group is the braid group $B_N$, while in three-dimensional space we obtain the permutation group $S_N$, leading to bosons and fermions. 
For particles moving on graphs, we can obtain a variety of exchange groups, dubbed graph braid groups.  
These have recently been analysed in some detail in 
\cite{QuantumStatistics_Tomasz,Universal_Tomasz,GeometricPresentation}
and appear to be the natural tool to study the exchange statistics of particles on wire networks, without reference to a two dimensional medium.   
Given that anyons can be braided while restricting their motion to planar graphs, the corresponding representations of $B_N$ must appear also as a representation of a planar graph braid group. 
For sufficiently connected planar graphs, graph braids satisfy the same relations as braids on the plane or surface.

However for less connected graphs, such as a trijunction, graph braiding affords more freedom and it has been conjectured ~\cite{Universal_Tomasz} that there may be braid statistics on a graph which do not exist in the plane.
In fact, for simple junctions, graph braid groups are free groups, allowing for arbitrary braid actions on the Hilbert space. This strongly suggests that more physical input is needed to pick specific graph braid representations.  
An important piece of information we can add is the fusion of the particles' topological charges. For anyons, this leads to the framework of anyon models, or more precisely unitary braided tensor categories. The braiding is constrained through the hexagon equations, which enforce compatibility of braiding and fusion. 
We now develop the basics of an analogous framework for particles on graphs.

\begin{figure}[t]
      \centering
 	\resizebox{0.5\linewidth}{!}
    { \includegraphics{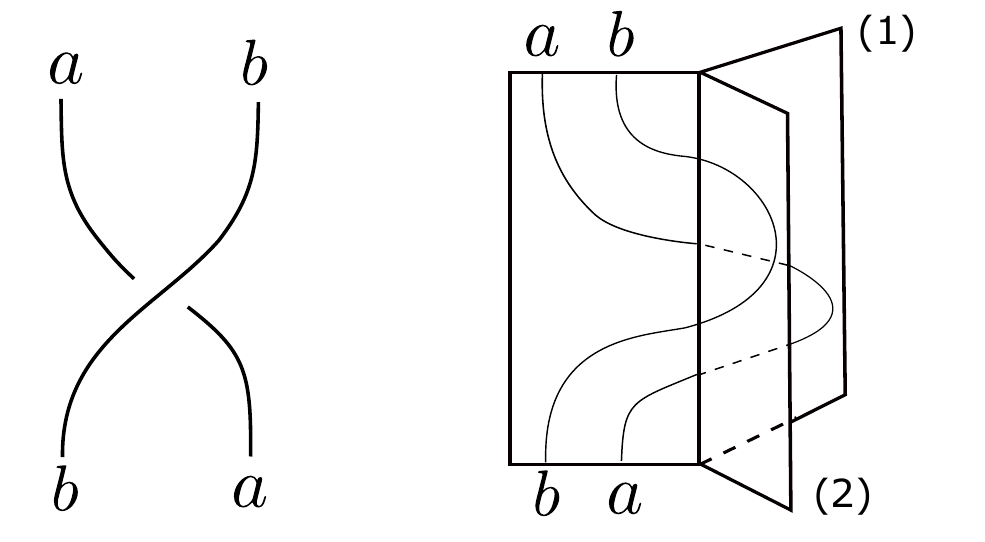}
	}
\caption{Diagrams for a simple two particle exchange on the plane, $\tau_1$ (left) and on a trijunction, $\sigma_{1}^{(1,2)}$ (right). }
\label{fig::DefGraphBraid}
\end{figure}

\paragraph{Quantum exchange statistics and graph braid groups.}
The $N$-strand graph braid group $B_{N}(\Gamma)$ of a graph $\Gamma$ is defined~\cite{Universal_Tomasz,QuantumStatistics_Tomasz,GeometricPresentation}~by $B_{N}(\Gamma)=\pi_{1}(C_{N}(\Gamma))$. Here $C_{N}(\Gamma)$ is the space of configurations of $N$ identical particles in distinct positions on $\Gamma$. 
For convenience we take the base point so that all particles are located on a single edge of the graph. 
A graph braid is then represented by the spacetime history where the particles start at their positions on this initial edge, are then transported to other edges and finally returned to the initial edge, possibly with the order of some of the particles changed. 
An example of a two particle exchange at a trijunction can be seen in Fig.~\ref{fig::DefGraphBraid}. Clearly each junction in the graph offers an opportunity to exchange particles in this way.   
An intuitive presentation of $B_N(\Gamma)$ based on 2-particle exchanges is given in \cite{GeometricPresentation}. 
This presentation has generators denoted by $\sigma_{j}^{(a_{1}, a_{2}, \dots, a_{j} , a_{j+1})}$,
where $a_{i}$ denotes the edge that the $i^{\mathrm{th}}$ particle away from the junction point is moved to during the graph braid.
For general graphs one makes this unambiguous by first choosing a unique path out to each edge. 
This is provided by the spanning tree of the graph.
The subscript $j$ denotes that after the action of $\sigma_{j}$ the particles return to the initial edge in the same ordering except that particle $j+1$ returns before particle $j$, so that these particles end up on the initial edge in reverse order.
Note that in exchanging particle $j$ with particle $j+1$, it is necessary to move all particles ahead of particle $j$ in order to get particle $j$ to a vertex, where it can then be exchanged. 
The inverse of a $\sigma_{j}$ generator is given by switching $a_{j}$ and $a_{j+1}$.
We can contrast these generators with the well known presentation of the planar braid group $B_N$ generated by $\tau_{i}$, which exchanges two neighbouring strands labelled $i$ and $i+1$, subject to the following relations,
\begin{equation}
		\tau_{i} \tau_{j} = \tau_{j} \tau_{i} \hspace{10pt} |j-i| >2  
		~~\mathrm{and}~~
		\tau_{i+1} \tau_{i} \tau_{i+1}  
		= \tau_{i} \tau_{i+1} \tau_{i} 
\end{equation}
We may call these relations the local commutativity (left) and Yang-Baxter relations (right). 
All the $\sigma_{j}^{(a_{1}, a_{2}, \dots, a_{j} , a_{j+1})}$ operators would correspond either to $\tau_j$ or to $\tau_{j}^{-1}$ if the particles were free to move in the plane, but when motion is restricted to the graph, it is necessary to keep track of the edges where the rest of the particles that are moved out of the way go to during the motion and as a result graph braid groups have multiple counterpart generators for $\tau_j$ with $j>1$. 

In general graph braid groups 
have fewer relations than the planar braid group in the following sense \cite{Universal_Tomasz, GeometricPresentation}. For some, but not all pairs of $\sigma_i$, $\sigma_j$ generators (depending on the upper indices) there are relations similar to local commutativity, these are called pseudocommutative relations and similarly for some  $\sigma_j$ generators there are relations analogous to the braid relation, called pseudobraid relations. 
The overall structure of graph braid groups is often quite simple. For example, if the graph has one vertex and $d$ edges, the graph braid group is isomorphic to a free group. Some of the $
\sigma_j$ generators can be eliminated, depending on $d$, but a number remain and those have no further relations. 
This means these generators can be represented by any unitary operators on the Hilbert space and more physical information will be needed to actually determine the effect of particle exchanges. 
We shall focus on $N=3$ particles on a trijunction.
This is one of the most familiar set ups for TQC, \cite{Alicea_Junction,ParafermionTrijunc,Topological_Josephson_MBS}.
The graph braid group is generated by,
\begin{equation}
\label{eq::TrijunctionGraphBraidGroup}
 B_{3}(\Gamma_{3}) = \langle \, \sigma_{1}^{(1,2)}, \quad  \sigma_{2}^{(2,1,2)}, 
 \quad \sigma_{2}^{(1,1,2)} \, \rangle.
\end{equation}
In this case there are no pseudocommutative relations and it can be verified graphically that there are no pseudobraid relations either, because the required path deformation would require two particles to occupy the vertex simultaneously, which is forbidden.  
Hence $B_{3}(\Gamma_{3})$ is a free group on three generators and we need more physical input, to constrain the unitary operators which implement exchanges.
To this end we introduce topological charges and fusion into the picture.

\begin{figure}[t]
     \centering
 	\resizebox{0.65\linewidth}{!}
   { \includegraphics{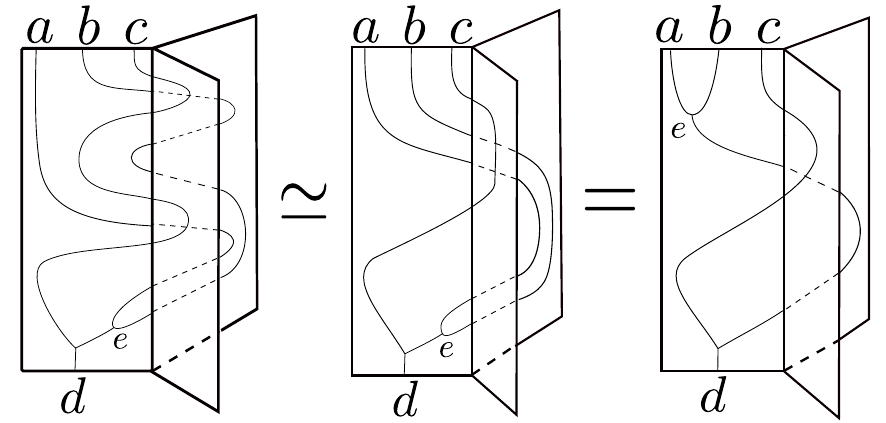}
	}
 \caption{Here we show an example of
  sliding a fusion vertex $e = a \times b$ through a graph braid.
}
\label{fig::FusionCommutesGraphBraiding}
\end{figure}

\paragraph{Graph anyon models.}
We now construct the elements of graph anyon models, analogous to the planar braiding of anyons, see e.g.~\cite{Kitaev2006,CFT_Anyon_Seiberg}. 
A more complete presentation will be given in \cite{Long_article_in_prep}.
Particles in a graph anyon model carry one of a finite set of topological charges $a,b,c,..$, there is  fusion of charges,
\begin{equation}
	a \times b = \sum_{c} N_{c}^{ab} c.
\end{equation}
The coefficient $N_{c}^{ab}\in \mathbb{Z}_{\ge 0}$ is the dimension of the fusion space $V^{ab}_{c}$ of ground states on a single edge, with two particles of charges $a$ and $b$ and with overall charge $c$.
Here, we will consider only multiplicity free models, so  $N^{ab}_{c}\in\{0,1\}$.
There is a unique vacuum charge $1$, such that $a \times 1 = 1 \times a = a$ for all $a$. 
Also each charge $a$ has a unique conjugate, $a \times \bar{a} = 1$.
We choose an orthonormal basis for each nontrivial fusion space $V^{ab}_{c}$. 
This choice introduces a gauge freedom $u^{ab}_{c}$, a unitary matrix of dimension $N^{ab}_{c}$ which changes the basis and leaves the physics unchanged. 
In the multiplicity free case, $u^{ab}_{c} \in  U(1)$.
We can form multiparticle states from tensor products of fusion spaces.  
This leads to two alternative bases for the three particle space of charges $a,b,c$ with total topological charge $d$, related by a change of basis whose matrix elements $[F_{d}^{abc}]_{e,f}$ are called the $F$-symbols;
\[
  \centering
  \begin{tikzpicture}[scale = 0.33]
		\node (0) at (-8.75, 1.5) {};
		\node (1) at (-7.75, 0.5) {};
		\node (2) at (-6.75, 1.5) {};
		\node (3) at (-6.75, -0.5) {};
		\node (4) at (-8.75, 2) {$a$};
		\node (5) at (-6.75, 2.13) {$b$};
		\node (6) at (-4.75, 2) {$c$};
		\node (7) at (-4.75, 1.5) {};
		\node (8) at (-6.75, -1.75) {};
		\node (9) at (-7.8, -0.25) {$e$};
		\node (10) at (-7.25, -1.75) {$d$};
		\node (11) at (0.5, 1.5) {};
		\node (13) at (3.0, 1.5) {};
		\node (14) at (3.0, -0.5) {};
		\node (15) at (0.5, 2) {$a$};
		\node (16) at (3.0, 2.13) {$b$};
		\node (17) at (5.0, 2) {$c$};
		\node (18) at (5.0, 1.5) {};
		\node (19) at (3.0, -1.75) {};
		\node (20) at (4.1, -0.25) {$f$};
		\node (21) at (2.5, -1.75) {$d$};
		\node (22) at (4.2, 0.6) {}; 
		\node (23) at (-2, 0) {$=\sum\limits_{f}\left[F^{abc}_d\right]_{ef}$};
		\draw (0.center) to (1.center);
		\draw (1.center) to (2.center);
		\draw (1.center) to (3.center);
		\draw (3.center) to (7.center);
		\draw (3.center) to (8.center);
		\draw (14.center) to (18.center);
		\draw (14.center) to (19.center);
		\draw (11.center) to (14.center);
	\draw (13.center) to (22.center);
\end{tikzpicture}
\]
The $F$-symbols are required to satisfy the \textit{pentagon equation}, which assures that the order of fusion can be consistently rearranged locally for systems with any number of particles~\cite{CFT_Anyon_Seiberg,Kitaev2006}.
The description of braiding in planar anyon models is implemented by a unitary operation $R$. 
Its effect on
$V^{ab}_{c}$ is given by the \textit{$R$-symbols} $R^{ab}_{c}$ which are $U(1)$ matrices. 
The compatibility of fusion and braiding is often phrased by saying that \textit{fusion commutes with braiding}. In spacetime diagrams it means that we can slide a particle worldline under or over a fusion or splitting vertex.
To make this consistent, the $R$-symbols must satisfy the \textit{hexagon equations}. 
For braiding on a graph, the usual hexagon equations are not valid, in fact fusion and braiding do not always commute. However, there are still particular processes where a continuous deformation of the particles' history leads to an exchange of a fusion with a braiding in time, see Fig.~\ref{fig::FusionCommutesGraphBraiding} for an example. 
We now define appropriate symbols satisfying graph hexagon equations which express this remaining consistency of fusion and braiding on a trijunction, 
\begin{equation}
\hspace{-4pt}
\rho(\sigma_{1}^{(1,2)}) := R , \quad  \rho(\sigma_{2}^{(2,1,2)})
	 := Q, \quad \rho(\sigma_{2}^{(1,1,2)}):= P .
\label{eq::TrijuncGraphMat}
\end{equation}
The action of $R$ on $V^{ab}_{c}$
is given by $R$-symbols, 
\[
 	\resizebox{0.41\linewidth}{!}
    { \includegraphics{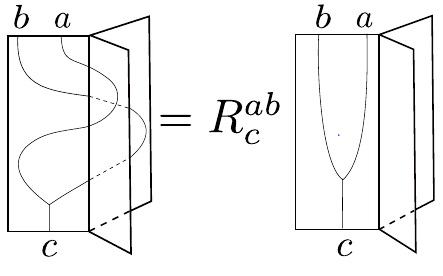}
	}
\]
Note these $R$-symbols aren't necessarily solutions of the planar hexagon equations. 
Similarly, the graphical representations of $P$ and $Q$, which exchange the second and third particles away from the vertex are,
\[
	\resizebox{1.00\linewidth}{!}
	{ \includegraphics{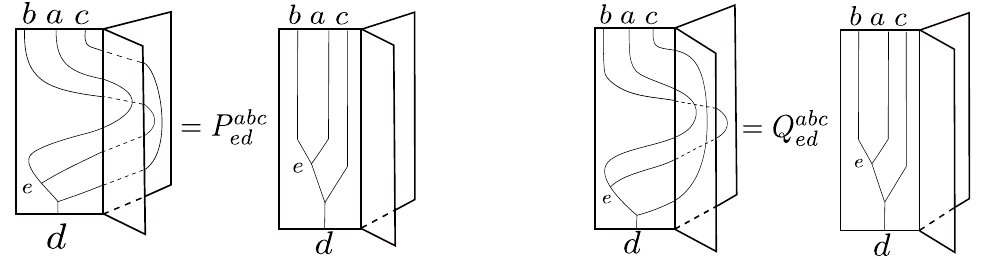}
	}
\]
Note that these braiding processes necessarily involve all three particles labeled $a$, $b$, $c$ and so we have introduced additional labels characterizing the full three particle state to label matrix elements of the  graph braid matrices. If we made $P$ and $Q$ only depend on $a,b$ and their fusion outcome, we would have $P=Q$, despite the fact they represent different generators in the graph braid group. 
As in the planar case, the $P$, $Q$, $R$ symbols are $U(1)$ matrices acting on the states of $V^{ab}_{c}$.
Gauge transformations have a similar effect on $P$, $Q$ and $R$, we have 
\begin{equation}
	\label{eq::GraphBraidGaugeTransform}
    R^{ab'}_{c} = \frac{u^{ba}_{c}}{u^{ab}_{c}} R^{ab}_{c},	\qquad
    W^{abc'}_{ed} = \frac{u^{ba}_{e}}{u^{ab}_{e}} \, W^{abc}_{ed},\quad \, W\in\{P,Q\}.
\end{equation}

\begin{figure*}
     \centering
 	\resizebox{0.72\linewidth}{!}
    { \includegraphics{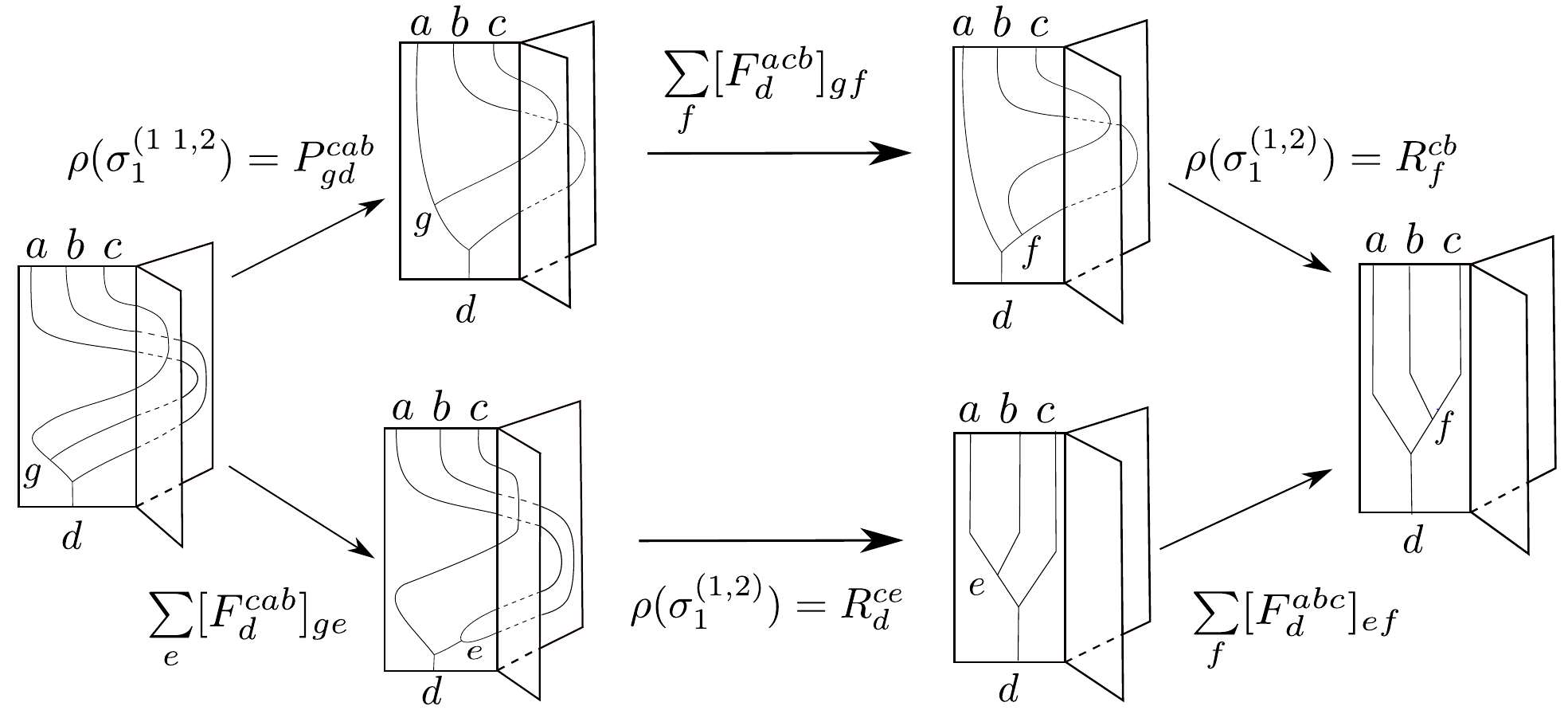}
	}
\caption{Here we show a hexagonal commutative diagram enforcing compatibility of fusion and graph braiding. In the bottom left state we have used the premise that fusion commutes with graph braiding, which we display in Fig.\ref{fig::FusionCommutesGraphBraiding}.}
\label{fig::HexagonOne}
\end{figure*}
We now consider compatibility of graph braiding and fusion. If a particle braids over two other particles with a given total charge (fusion channel), the process must involve two individual exchanges, see Fig.~\ref{fig::FusionCommutesGraphBraiding}. 
Here, the two exchanges are such that we can \emph{slide} a fusion vertex through a graph braid. 
This implies in particular that the two particles with the joint charge $e$ must go to the same edge during the braid process. 
Otherwise the move pushing the splitting vertex upward would be blocked at the graph's vertex  and it would not pass under the worldline of charge $c$. 

Adapting the notation from the $\sigma$ presentation, we can write an equation for the diagram identity in Fig.~\ref{fig::FusionCommutesGraphBraiding} as follows,
\begin{equation}
	\sigma_{2}^{(1_{b},1_{a},2_{c})} \circ \sigma_{1}^{(1_{b},2_{c})} 
	= \sigma_{1}^{(1_{b \times a},2_{c})} = \sigma_{1}^{(1_{e},2_{c})} .
\end{equation}
We can now construct our consistency equations for graph braiding and fusion: the graph hexagon equations. We note that we can connect the two sides of the identity in Fig.~\ref{fig::FusionCommutesGraphBraiding} by a series of $F$-moves and exchanges and we then require this combination of moves to be the identity. This leads to the hexagonal commutative diagram in Fig.~\ref{fig::HexagonOne}. 
We can similarly obtain another such equation in a situation where the joined particles move to the other plane.  
This leads to the equations;
\begin{eqnarray}
	\label{eq::GraphHexagonEquations}
	 P^{cab}_{gd} \,  \left[F^{acb}_{d}\right]_{gf}
	  \,  R^{cb}_{f}  
	 &=& 	
	 \sum_{e} \left[F^{cab}_{d}\right]_{ge} \,	 
	 R^{ce}_{d}  \,  \left[F^{abc}_{d}\right]_{ef},  
	\\
	(Q^{acb}_{gd})^{*} \,   \left[F^{acb}_{d}\right]_{gf}  
	\, (R^{bc}_{f})^{*}
	 & =&  \sum_{e} \left[F^{cab}_{d}\right]_{ge} \,	
	(R^{ec}_{d})^{*}  \, \left[F^{abc}_{d}\right]_{ef}.  
	\nonumber 
\end{eqnarray}
Of course similar equations can also be derived starting from the inverse braids. 
These are equivalent to the ones given and involve $P^{-1}$ and $Q$.
We can also check that the equations are consistent with simple physical requirements
such as $P^{1ab}_{ac}=P^{a1b}_{ac}=1$ and $P^{ab1}_{cc}=R^{ab}_{c}$, and similarly for $Q$.

\paragraph{Solutions of the graph hexagon equations.}
First of all, we note that, when $P^{abc}_{ed}=Q^{abc}_{ed}=R^{ab}_{e}$, for all $a,b,c,d,e,$ the graph hexagons reduce to the usual hexagon equations for planar systems, and the corresponding graph braid group representations are necessarily representations of $B_N$. 
Hence any planar anyon model immediately provides solutions to these equations, although often further solutions exist. 
We now consider some simple fusion models to illustrate what else is possible.  
Since the graph braid matrices and $F$-matrices are invertible, we can immediately use the hexagon equations (\ref{eq::GraphHexagonEquations}) to express the
$P$ and $Q$ symbols in terms of the $R$ symbols. 
This means we only need to supply the $F$-symbols and find the $R^{ab}_{c}$ to fix all symbols. 
Notice that we usually get multiple expressions for the same $P$ or $Q$ symbol, as the index $f$ varies. This will restrict the possible values for the $R^{ab}_{c}$.  
However, in Abelian fusion models, given $F$, the $R^{ab}_{c}$ are not restricted and can be freely chosen. In such models, the charges $a,b,..$ are elements of a finite Abelian group $G$ and the fusion corresponds to group multiplication, giving a unique outcome for each fusion. In this case the label $f$ in Eq. (\ref{eq::GraphHexagonEquations}) is fixed as the unique fusion of $b$ and $c$. 
 Hence a choice of $R^{ab}_{a\times b}$ just fixes the $P$ and $Q$ symbols. T
 here are no requirements on $R$, apart from $R^{ab}_{c}\in U(1)$ and $R^{a1}_{a}=R^{1a}_{a}=1$.
This already gives us many examples which do not satisfy the planar hexagon equations. For example when $G=\ZZ_{M}$, the $F$-symbols satisfying the pentagon equations are given by $3$-cocycle $\omega$ in the group cohomology of $G$
\cite{CFT_Anyon_Seiberg}.
The solutions $R^{ab}_{c}$ to the planar hexagons for $\mathbb{Z}^{\omega}_{M}$ with trivial $\omega$ are required to form a non-degenerate symmetric bicharacter $\chi(a,b)$, but no such requirement is needed on a graph. 
Perhaps more interestingly, there is often no nontrivial solution to the planar hexagon. 
This occurs e.g. when $M$ is odd and $\omega(a,b,c)$ is cohomologically nontrivial~\cite{BondersonInterfer}. 
Nevertheless there is a solution for any choice of the $R^{a,b}_{a\times b}$ on the trijunction and so we can graph braid particles that do not permit planar braiding. 
Overall, for any group $G$ we obtain a $(|G|-1)^2$ parameter family of solutions for any fixed choice of $F$-symbols. Some of these will be related through the gauge freedom Eq. (\ref{eq::GraphBraidGaugeTransform}). 
However, since the symbols $R^{aa}_{a\times a}$ and the products $R^{ab}_{a\times b}R^{ba}_{b\times a}$ are independent and gauge invariant for all $a,b\neq 1$, we always have at least $|G|(|G|-1)/2$ physical parameters.
For theories with non Abelian braiding we must equate the expressions for $P$ and $Q$ which come from different choices of $f$ in Eq.~(\ref{eq::GraphHexagonEquations}). 
This then yields equations purely for $R^{ab}_{c}$.
For example for the Fibonacci model, which has a single nontrivial charge $\tau$ with $\tau\times\tau=1+\tau$, we recover the known planar values for the $R$-symbols as the only solutions, and $P^{abc}_{ed}=Q^{abc}_{ed}=R^{ab}_{e}$.  
Another important example is the Ising theory, this model is directly relevant for topological memories based on quantum wires that host Majorana modes ~\cite{Sarma_Network_TQC,Alicea_Junction}.
The Ising theory is an example of a larger family known as the Tambara-Yamagami models ~\cite{TamYama}. We have charges $g_{i}$ forming a finite Abelian group, $G$ and a single additional charge $\sigma$ such that $\sigma\times \sigma =\sum_{i} g_{i}$ and $g_i \times \sigma=\sigma$. 
The case $G=\ZZ_2$ is the Ising model with $(1,\sigma,\psi) \equiv(g_0,\sigma,g_1)$ being the more usual notation. 
All these fusion rules have known $F$-symbols, but in the plane they allow no solutions to the hexagon equations unless $G = (\ZZ_{2})^n$ for some $n$~\cite{TM_no_Hex}.
We find there is similarly no solution to the graph hexagon equations unless $G=\ZZ_{2}^{n}$, ~\cite{Long_article_in_prep}.
However, when $G=\ZZ_{2}^{n}$ graph braid solutions without a counterpart in the plane exist. We state some results for the Ising model.
There are two possible solutions of the pentagon equations but is turns out that these lead to the same set of $R$-symbols for the graph hexagon equations, namely
\begin{equation}
\label{eq::GraphIsing}
    R^{\sigma \sigma}_{1} = \pm i R^{\sigma \sigma}_{\psi}
    ~~~
    R^{\sigma \psi}_{\sigma} = R^{\psi \sigma}_{\sigma} = \pm i
    ~~~
    R^{\psi \psi}_{1}\in U(1)
\end{equation}
These solutions are inequivalent under gauge transformations which fix the $F$-symbols. 
The free parameters, $R^{\psi \psi}_{1}$ and $R^{\sigma \sigma}_{1}$, are gauge invariant. 
The planar Ising solutions appear for $R^{\psi \psi}_{1}=-1$ and $R^{\sigma\sigma}_{1}=e^{\pm i (2k+1)\pi/8}$, where $k\in\{0,1,2,3\}$\footnote{Note that on the plane, the cases $k=0$ and $k=3$ occur for one set of $F$-symbols while the remaining cases occur for the other set. Here, all options occur for either choice of $F$-symbols.}. 
The $P$ and $Q$ symbols are now easily obtained, and given in the supplementary material. 
They depend on the chosen pentagon solution.

\paragraph{Discussion and Outlook}
We have presented only the basic features of braiding and fusion on graphs here, stressing the differences with planar systems.  In a preprint in preparation, \cite{Long_article_in_prep} we consider general networks with more and higher valence vertices and loops, and with more particles braiding, as well as further implications for TQC. As a taster, we work out the case of three particles on a $4$-valent junction in the supplementary material, focusing on the effect of the pseudobraid relation which first appears there.  
Many features presented here persist generally, for example using compatibility of braiding and fusion we can always express generators $\sigma_{j}$ for exchanges of particles further from the vertex in terms of exchanges $\sigma_{j'<j}$ of those closer to the vertex (though details depend on the valence).
An interesting question is to find a generating set for all compatibility constraints for braiding and fusion for any number of particles on any graph. This may be most elegantly addressed in a categorical setting where it would lead to an analogue of the MacLane coherence theorem~\cite{maclane1963natural}.
Another natural question is under what circumstances the coherence is strong enough to yield Ocneanu rigidity\cite{Kitaev2006,etingof2005fusion}, meaning that the set of solutions modulo gauge is finite. This clearly does not hold for the trijunction, but this property returns for networks with loops. 

\paragraph{Acknowledgements.}
The authors would like to extend special gratitude to Dr. Tomasz Maci{\k{a}}{\.z}ek for introducing the authors to this topic.
J.K.S. acknowledge financial support from Science Foundation Ireland through Principal Investigator Awards 12/IA/1697 and 16/IA/4524.
A.C. was supported through IRC Government of Ireland Postgraduate Scholarship GOIPG/2016/722.
The authors acknowledge financial support from the Helbronn Institute, which facilitated a workshop on this topic.
A.C. would like to thank Dr. Ian Jubb for helpful discussions.

\bibliographystyle{unsrt}
\bibliography{Anyon_Bibliography}
 \appendix
%
\section{Supplementary material.}

\paragraph{Tetrajunction.}
The tetrajunction $\Gamma_4$ is one of the simplest graphs for which the graph braid group contains a pseudobraid relation.
Additionally we can observe that $B_{3}(\Gamma_{4})$ contains three sub-trijunctions, coming from assigning the particles to an initial edge and then choosing one of the three choices of pairs of the remaining three edges to exchange them. 
The graph braid group $B_{3}(\Gamma_{4})$ is a free group of rank $11$, but there are $12$ elements in the $\sigma$ presentation. One of these can be eliminated by means of the pseudobraid relation~\cite{GeometricPresentation}. 
The matrices representing the generators of $B_{3}(\Gamma_{4})$ can be written
\begin{eqnarray}
    &\rho(\sigma_{1}^{(1,2)}):= X,  \qquad \rho(\sigma_{2}^{(a_{1},1,2)}) 
    = X_{a_{1}}, \nonumber \\
    &\rho(\sigma_{1}^{(2,3)}):= Y, \qquad \rho(\sigma_{2}^{(a_{1},2,3)}) 
    = Y_{a_{1}},  \\
    & \rho(\sigma_{1}^{(1,3)}):= Z, \qquad \rho(\sigma_{2}^{(a_{1},1,3)}) 
    = Z_{a_{1}} \nonumber.
\end{eqnarray}
Here, $a_{1} \in \{1,2,3\}$ labels the edge that the particle closest to the junction point goes to during the graph braid. 
This notation highlights the trijunction subgroups. Referring to the notation used for the trijunction in Eq.~(\ref{eq::TrijuncGraphMat}), we see that the $R$-matrices (given by exchanging the two particles closest to the junction point) for each sub trijunction occur in the first column above and are now labeled $X$, $Y$ and $Z$ for the three trijunctions. 
The $P$ and $Q$ graph braid matrices appear in the second column and for example the trijunction which utilizes edges $1$ and $2$ has 
$R\equiv X, \, P\equiv X_1$ and $ Q\equiv X_2$. Similarly $(Y, Y_2, Y_3)$ and $(Z, Z_1,Z_3)$ also generate trijunction subgroups. The generators $X_3$, $Y_1$ and $Z_2$ utilize all edges and have no counterpart on a trijunction. 
Consistency of braiding and fusion now comes down to graph hexagon equations similar to Eq.~(\ref{eq::GraphHexagonEquations}) on each sub trijunction, yielding $6$ independent sets of equations. No hexagon equations exist for the generators that involve all three edges. 
If one tries to commute a fusion vertex through a graph braid involving one of these generators, the fusion vertex will get blocked on the junction point.

Since we just have three independent copies of the graph hexagons for the trijunction, they can be solved as before. However, one can make independent choices of solutions for each set of trijunction hexagon equations. 
E.g. in the case of the Ising fusion rules, one could have, say, $X^{\sigma \psi}_{\sigma} = +i \, $  and $Y^{\sigma \psi}_{\sigma} = Z^{\sigma \psi}_{\sigma} = -i$. 
Similarly, for the Fibonacci model, which only allows the usual planar solutions on the trijunction, we can now choose solutions of different chirality on the subjunctions, which yields non-planar solutions for this model on the tetrajunction.  
The generators $X_{3}$, $Y_{1}$ and $Z_{2}$, which use all edges, occur in the representation of the pseudobraid relation,
 \begin{equation}
 	\label{eq::PseudoBraid}
    \sigma_{2}^{(1,2,3)} \, \sigma_{1}^{(1,3)} \, \sigma_{2}^{(3,1,2)} 
    = \sigma_{1}^{(1,2)} \, \sigma_{2}^{(2,1,3)} \, \sigma_{1}^{(2,3)}.
\end{equation}
This is a graph braiding analogue of a Yang-Baxter equation.
One may write $6$ such relations for different permutations of $(1,2,3)$, but only one is independent. 
 \begin{figure}[t!]
    \centering
   	\resizebox{0.60\linewidth}{!}
      { \includegraphics{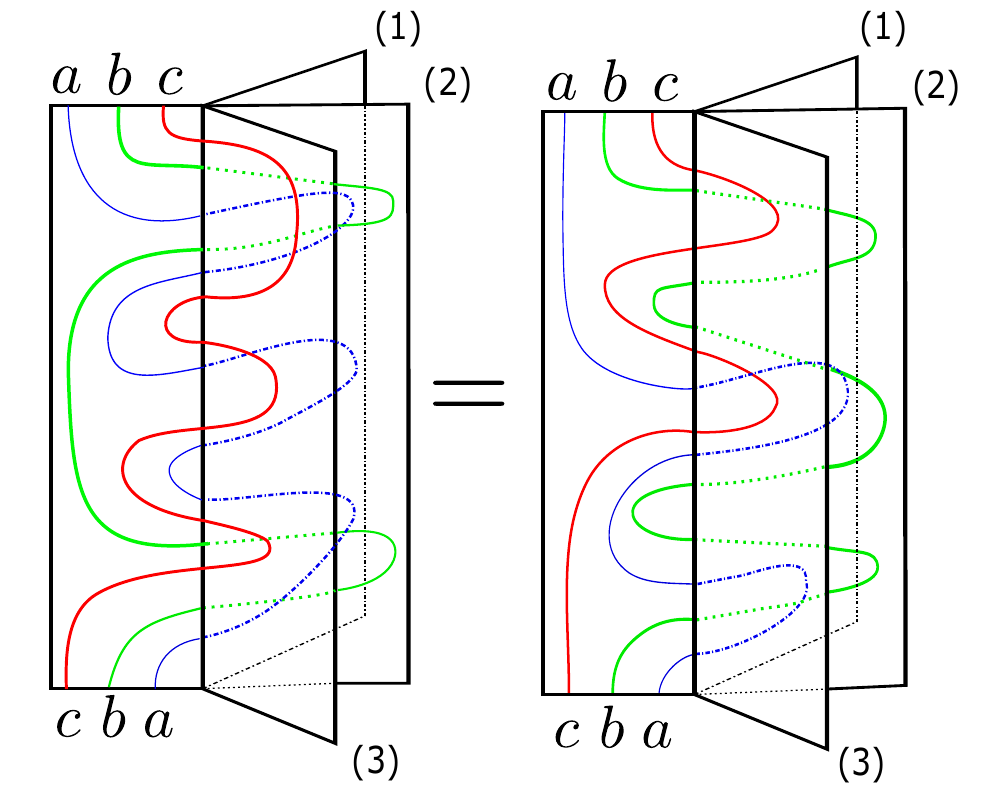}
  	  }
	  \caption{(Colour online) Graphical representation of the pseudobraid relation on the tetrajunction. The left diagram corresponds to the left hand side of Eq.(\ref{eq::PseudoBraid}), with composition in the equation going vertically in the diagram. The edge assignment for the particles stays fixed throughout the composition on both sides of the equality. }
 \label{fig::PseudoBraid}
 \end{figure}
We can write the pseudobraid relation in terms of the $X$, $Y$ and $Z$ symbols by introducing fusion trees at the bottom of the diagrams in Fig.\ref{fig::PseudoBraid}. The equality then induces a dodecagonal commutative diagram of $F$-moves and exchanges, similarly to how the equality expressing the compatibility of fusion and braiding induced the graph braiding hexagon equations in Fig.\ref{fig::HexagonOne}.
This finally gives the following equation,
  \begin{equation}
 	\begin{split}
 	&  {Y_{1}}^{cba}_{fd}
		\, \sum  \limits_{e,g} 
 		\left[F^{bca}_{d}\right]_{fg}	\,
 	Z^{ca}_{g}
 	\, 	\left[(F^{bac}_{d})^{-1}\right]_{ge}
 	\,{X_{3}}^{bac}_{ed}
 	\,	\left[F^{abc}_{d}\right]_{ef}	\\
 	&=    Y^{cb}_{f}
	\sum \limits_{e,g} 
	\, 	\left[F^{cba}_{d}\right]_{fe}
 	\,X^{ba}_{e} \,
 	\left[(F^{cab}_{d})^{-1}\right]_{eg}
 	\, {Z_{2}}^{cab}_{gd} \,
 	 \left[F^{acb}_{d}\right]_{gf}
 	 \\
 	\end{split}
 	\label{eq::pseudobraidsymbols}
 \end{equation}
This equation simply allows us to fix the $Y_1$ symbols in terms of the other symbols. 
There is never a conflict with the graph hexagons for the trijunctions, as they never involve $Y_1$.  
Of course, by choosing a different pseudobraid relation, we could choose to eliminate the $X_3$ or $Z_2$ symbols if we prefer. 
Note that, although we can always eliminate one of the families of symbols that involve all three edges, the other two families are free parameters, as they are not constrained by any further equations. 
This means that for example for Abelian fusion rules governed by a group $G$, we end up having an extra $2(|G-1)^3$ free parameters in addition to the $3(|G|-1)^2$ parameters coming from the trijunctions. 
The situation is similar for non-Abelian models -- we have free parameters for all of these, although the actual parameter counting is a little more complicated. 
For example, the Fibonacci model has $3$ free $X_3$ symbols and $3$ free $Z_2$ symbols, even though it allows no free parameters at all on the trijunction.  

Eq.~(\ref{eq::pseudobraidsymbols}) simplifies for Abelian fusion rules. For these,  $\left[F^{abc}_{a\times b\times c}\right]_{a\times b,b\times c}=\omega(a,b,c)$, where $\omega$ is a $U(1)$ valued group $3$-cocycle. We find
\begin{equation}
	{Y_{1}}^{cba}_{fd} \, c_{a}(b,c) \,{X_{3}}^{bac}_{ed} \, Z^{ca}_{g} 
	\, =\, Y^{cb}_{f} \, c_{a}(c,b) \, X^{ba}_{e} \, {Z_{2}}^{cab}_{gd}.
\end{equation}
Where $c_{a}(b,c)$ is the Slant product,
\begin{equation}
	c_{a}(b,c) := (i_{a}\omega)(b,c) = \frac{\omega(a,b,c) \omega(b,c,a)}{\omega(b,a,c)}
\end{equation}
Often, we can take $c_{a}(b,c)=c_{a}(c,b)$, this happens e.g. for all $\mathbb{Z}^{\omega}_{M}$ anyon models. In that case, the Abelian pseudobraid equation does not depend on the $F$-symbols.

Finally we note that we also have new types of gauge invariant parameters on a tetrajunction -- for example, the gauge invariant quantities  
$W^{ab} (W')^{ba}$, where $W = \{X,Y,Z\}$ and $W \neq W'$, appear in addition to the parameters $W^{aa}$ and $W^{ab}W^{ba}$ which come from the  sub-trijunctions.

\paragraph{Ising solution.}
In this section we will solve the graph hexagon equations for the Ising fusion rules on a trijunction.
The planar solutions to the pentagon equations and hexagon equations for these fusion rules can be found in ~\cite{pachos_2012,Kitaev2006}.
There are two non-equivalent sets of $F$-symbols, distinguished by the value of the Frobenius-Schur indicator $\nu=\pm 1$. Most $F$-symbols equal 1. 
The nontrivial $F$-symbols are $[F^{\sigma \psi \sigma}_{\psi}]=[F^{\psi \sigma \psi}_{\sigma}]= -1$ and 
\begin{equation}
    \left[F^{\sigma \sigma \sigma}_{\sigma}
	\right]_{ef} = 
	\frac{\nu}{\sqrt{2}}  \, F(e,f) = \frac{\nu}{\sqrt{2}}
    \begin{pmatrix}
    1 & 1\\
    1 & -1\\
    \end{pmatrix} 
\end{equation}
where $e,f \in \{1,\psi\}$. 
We first consider Eq.~(\ref{eq::GraphHexagonEquations}) with $a=b=c=d=\sigma$. 
Focusing on the equation for $P^{\sigma \sigma \sigma}_{1 \sigma}$ and substituting the $F$-symbols we find that
\begin{equation}
    P^{\sigma\sigma \sigma}_{1\sigma} = 
    \, \frac{\nu}{\sqrt{2}} \,  
    (R^{\sigma \sigma}_{f})^{*} F^{*}(1,f) \, \sum_{e}F(1,e) \,
    R^{\sigma e}_{\sigma} \, F(e,f). 
\end{equation}
We equate the expression for $P^{\sigma \sigma \sigma}_{1 \sigma}$ with $f=1$ to that with $f=\psi$ to get
\begin{eqnarray}
    P^{\sigma \sigma \sigma}_{1 \sigma}
    &=& \frac{\nu}{\sqrt{2}} \, (R^{\sigma \sigma}_{1})^{*} 
    \,
    \Big( R^{\sigma 1}_{\sigma}+R^{\sigma \psi}_{\sigma} 
    \Big) \nonumber
	\\
	&=& \frac{\nu}{\sqrt{2}} \, (R^{\sigma \sigma}_{\psi})^{*} 
	\,\Big( R^{\sigma 1}_{\sigma} - R^{\sigma \psi}_{\sigma} 
    \Big).
	\label{eq::Psigsig1}
\end{eqnarray}
Equating the two expressions we get
\begin{equation}
	(R^{\sigma \sigma}_{1})^{*} \Big( 1+ R^{\sigma \psi}_{\sigma} \Big) 		
   	=  (R^{\sigma \sigma}_{\psi})^{*} \Big( 1 - 
   	R^{\sigma \psi}_{\sigma} \Big).
	\label{eq::Pg1fpaper}	 
\end{equation}
Which gives us an expression constraining the $R$ symbols directly. 
Notice that this does not depend on $\nu$, and neither do other equations for the $R$ symbols, so the solutions for $R$ will not detect the dependence on the Frobenius Schur indicator. 
\\
Additionally we can take the Hermitian adjoint of Eq.(\ref{eq::Psigsig1}) to get an expression for 
$(P^{\sigma \sigma \sigma}_{1 \sigma})^{-1}$ and then imposing
$P^{\sigma \sigma \sigma}_{1 \sigma} \times (P^{\sigma \sigma \sigma}_{1 \sigma})^{-1} = 1$
with $f= 1$ we get,
\begin{eqnarray}
	 1&=&\frac{\nu^2}{2} \, (R^{\sigma \sigma}_{1})^{*} \,
	(1 + R^{\sigma \psi}_{\sigma} ) \, R^{\sigma \sigma}_{1}  \, 
	 (1 + (R^{\sigma \psi}_{\sigma})^{*}) \nonumber \\ 
     &=& 1+ \frac{1}{2}\left(
     R^{\sigma \psi}_{\sigma}+(R^{\sigma\psi}_{\sigma})^{*}\right)
\end{eqnarray}
and hence $ (R^{\sigma \psi}_{\sigma})^{*}= - R^{\sigma \psi}_{\sigma}$, which yields
$R^{\sigma \psi}_{\sigma} = \pm i$.
\\
Substituing this back into Eq.~(\ref{eq::Pg1fpaper}), 
we find that
\begin{equation}
	R^{\sigma \sigma}_{1} = R^{\sigma \psi }_{\sigma} \, 
	R^{\sigma \sigma}_{\psi}= \pm i R^{\sigma \sigma}_{\psi}.
\end{equation}
By equating the expressions for $Q^{\sigma \sigma \sigma}_{1 \sigma}$ with $f=1$ and $f=\psi$ we find
\begin{equation}
\label{eq::Qexp}
	(R^{\sigma \sigma}_{1})^{*}(1 + R^{\psi \sigma }_{\sigma}) 		
	=(R^{\sigma \sigma}_{\psi})^{*}(1 - R^{ \psi\sigma }_{\sigma}).
\end{equation}
Comparing Eq.(\ref{eq::Qexp}) with Eq.(\ref{eq::Pg1fpaper}) implies $R^{\sigma \psi}_{\sigma} = R^{\psi \sigma}_{\sigma}$, as for the planar solution to the Ising model.
\\
We can tabulate the resulting values for $P$ and $Q$. 
The value for  $a=b=c=d = \psi$
\begin{eqnarray}
	&& P^{\psi \psi \psi}_{1 \psi} =  Q^{\psi \psi \psi}_{1 \psi} 
	= (R^{\psi \psi}_{1})^{*}.
\end{eqnarray}
Two particles are $\psi$ and one particle is $\sigma$;
\begin{eqnarray}
	&& P^{\sigma \psi \psi}_{\sigma \sigma}= Q^{\psi \sigma \psi}_{\sigma\sigma} 
	= \pm i, 
	\nonumber \\
	&& P^{\psi \sigma \psi}_{ \sigma \sigma}=-Q^{ \sigma \psi \psi }_{\sigma \sigma} 
	= \pm i \,  (R^{\psi \psi}_{1})^{*},
	 \\
	&& P^{\psi\psi\sigma}_{1\sigma} = Q^{\psi\psi \sigma}_{1\sigma}=- 1.
	\nonumber
\end{eqnarray}
Two particles are $\sigma$, one particle is $\psi$ and the total topological charge $d = \psi$;
\begin{eqnarray}
	&& P^{\psi \sigma \sigma}_{\sigma \psi} = Q^{\psi \sigma \sigma}_{\sigma \psi} = \pm i, 
	\nonumber \\
	&& P^{\sigma \sigma \psi}_{1 \psi} =  Q^{\sigma \sigma \psi}_{1 \psi} = R^{\sigma \sigma}_{1}, 	
 \\
	&& P^{\sigma \psi \sigma}_{\sigma \psi} = Q^{\sigma \psi \sigma}_{\sigma \psi} = \pm i.  
	\nonumber
\end{eqnarray}
Two particles are $\sigma$, one particle is $\psi$ and the total topological charge $d = 1$;
\begin{eqnarray}
    &&	P^{\psi \sigma \sigma}_{\sigma 1} =  Q^{\sigma \psi \sigma}_{\sigma 1}
    = \mp i R^{\psi \psi}_{1},  \nonumber \\
    &&	P^{\sigma \psi \sigma}_{\sigma 1} =	Q^{\psi \sigma \sigma}_{\sigma 1} 
    = \pm i,  \\
    && P^{\sigma \sigma \psi}_{\psi 1} =  Q^{\sigma \sigma \psi}_{\psi 1} 
    = \mp i R^{\sigma \sigma}_{1}. \nonumber
\end{eqnarray}
The final configuration has $a=b=c=d=\sigma$;
\begin{eqnarray}
	&& P^{\sigma \sigma \sigma }_{1 \sigma} = Q^{\sigma \sigma \sigma }_{1 \sigma} 
	= \nu \, e^{\frac{\pm \pi i }{4}} \, (R^{\sigma \sigma}_{1})^{*}, 
	\nonumber \\
	&& P^{\sigma \sigma \sigma }_{\psi \sigma} = 
	Q^{\sigma \sigma\sigma }_{\psi \sigma} = \nu \,
	e^{\frac{\mp \pi i }{4}} \,  (R^{\sigma \sigma}_{1})^{*}.   
\end{eqnarray}
The corresponding values for $Q^{-1}$ and $P^{-1}$ are given by Hermitian adjoint. One may check by direct verification that all braid hexagon equations are now satisfied for any choice of $R^{\sigma\sigma}_{1}$ and $R^{\psi\psi}_{1}$ in $U(1)$ and for both choices of $\nu$ and of the signs. 
It is interesting to observe that we have $P\neq Q$ whenever $R^{\psi\psi}_{1}\neq -1$. In other words $P\neq Q$ unless $\psi$ is a fermion. 
Nevertheless, even if $R^{\psi\psi}_{1}= -1$, the solution is usually not planar, as $R^{\sigma\sigma}_{1}$ is still a free parameter. 

\end{document}